# Confined optical phonon modes in polar tetrapod nanocrystals detected by resonant inelastic light scattering

*Roman Krahne[a*], Gerwin Chilla[b], Christian Schűller[b+], Luigi Carbone[a], Stefan Kudera[a] Gianandrea Mannarini[a], Liberato Manna[a], Detlef Heitmann[b], Roberto Cingolani[a],*

[a]National Nanotechnology Laboratory of CNR-INFM c/o Istituto Superiore Universitario di Formazione Interdisciplinare, Università di Lecce, Via per Arnesano, 73100 Lecce, Italy

[b]Institute of Applied Physics, University of Hamburg, 20355 Hamburg, Germany

[+] present address: Institute of Experimental and Applied Physics, University of Regensburg, 93040 Regensburg, Germany

roman.krahne@unile.it

*Corresponding author:

roman.krahne@unile.it; tel. +39 0832 298 209; fax +39 0832 298 238

**Abstract**

We investigated CdTe nanocrystal tetrapods of different sizes by resonant inelastic light scattering at room temperature and under cryogenic conditions. We observe a strongly resonant behavior of the phonon scattering with the excitonic structure of the tetrapods. Under resonant conditions we detect a set of phonon modes that can be understood as confined longitudinal-optical phonons, surface-optical phonons, and transverse-optical phonons in a nanowire picture.

**Keywords:** nanocrystals, quantum wires, phonons, resonant Raman scattering

Recently, chemical synthesis has made remarkable advances in controlling the shape of colloidal nanocrystals leading to spheres, [1, 2] rods,[3-5] and branched nanostructures.[6, 7] The tetrapod (TP) represents an intriguing nanostructure where four nanorods branch out at tetrahedral angles from a central region.[6] The specific geometry of the tetrapods has already demonstrated new properties in charge transport [8] and photoluminescence [9].

Optical phonons in confined nanostructures have been studied experimentally and theoretically in a variety of semiconductor nanocrystals such as spheres [10, 11] and rods[12-16]. In nanocrystal dots the general picture is that the phonons couple to the optical excitations via the Froehlich interaction [11, 17-20]. In spherical nanocrystals the confinement leads to a redshift and broadening of the LO phonon mode[21] and to the observation of surface optical (SO) phonons[10, 22]. Nanorods instead can be regarded as nanowaveguides, and are an especially interesting system for the study of phonons because of their uniaxial anisotropy both in shape and in crystal lattice[5]. An experimental and theoretical study of SO phonons in nanorods has shown that the SO phonon excitation depends on the nanowire shape[16, 23], and Raman scattering experiments on nanorods have shown splittings of the phonon modes that originate from the lateral confinement[12-15]. In this letter we report the observation of confined phonons by resonant Raman scattering in tetrapod-shaped nanocrystals. We employ a nanowire picture to interpret the TP data (since TPs consists of joined nanorods) and find good agreement with the predicted splittings of the LO and TO phonons, as well as for the energy of the SO phonon.

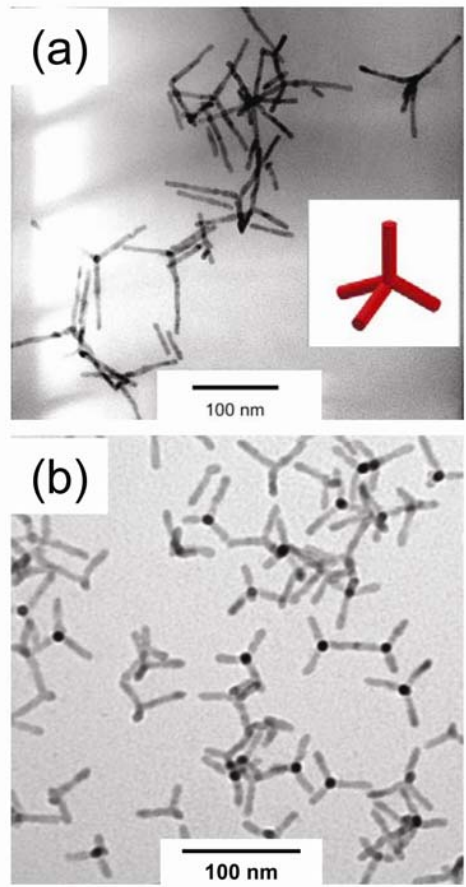

**Figure 1:** TEM images of CdTe tetrapods with different size; (a) large TPs with 12 nm / 80 nm arm diameter and length; (b) small TPs with 7 nm / 30 nm arm diameter and length. The inset shows a schematic drawing of a tetrapod.

CdTe TPs of different sizes were fabricated by chemical synthesis as reported in Ref.[6, 24] and dissolved in a solvent, typically chloroform. Figure 1 shows transmission electron microscope (TEM) images of two TP samples: (a) T1 with large and (b) T2 with small arm diameter and length. We investigated a series of tetrapod samples with arm diameter and length ranging from 5-12 nm and 18-80 nm, respectively. The solution containing the nanocrystals was drop-casted onto the surface of a silicon substrate, and the solvent was allowed to evaporate under soft nitrogen flow. The samples were then mounted in an optical cryostat, and the Raman experiments were performed using a tunable Ti-Sa Laser (700-850 nm). The excitation light (laser power 30 mW) was focused onto the samples on a spot of 50 micrometer diameter, and the signal was collected by an achromatic lens and detected by a triple Raman spectrometer (DILOR XY) and a CCD camera.

Figure 2a shows resonant Raman spectra of large tetrapods (arm diameter = 12 nm and length = 80 nm, see Fig. 1a) at room temperature. The laser excitation energy was varied from 1.678 eV to 1.755 eV and the spectra were normalized with respect to the amplitude of the Si substrate phonon. The resonant behaviour of the LO phonon, and of its second order scattering mode 2LO, is clearly observed and the extracted amplitudes of the excitations are reported in Fig. 2c. A spectrum at maximum resonance is shown in Fig. 2b and the positions of the observed peaks are evaluated by Lorentz fits. We find the TP LO phonon at 168 cm$^{-1}$ and its second order scattering at 336 cm$^{-1}$. Another weak excitation can be identified at 307 cm$^{-1}$, which most likely results from higher order scattering. Interestingly, the energy of this mode matches almost exactly the sum of the measured LO phonon and the bulk value of the TO phonon[25] (168 cm$^{-1}$+140 cm$^{-1}$=308 cm$^{-1}$).

We now focus on the resonant behaviour of the phonon excitations and plot the amplitude of the phonons with respect to the laser excitation energy in Fig. 2c. For comparison, we show also the emission and absorption of this sample (taken from Ref [9]). We find the resonance maximum of the LO phonon at 1.715 eV, slightly blue-shifted with respect to the energy of the first absorption peak. This is consistent with the expectation of a strong resonance maximum at one phonon energy above the exciton energy for the so called outgoing resonance. The outgoing resonance, which should occur at exactly the exciton ground state energy plus one phonon energy, has been shown for bulk samples to be much stronger than the incoming resonance [26-28]. We think that due to broadening the incoming and outgoing resonance maxima, which should differ by one LO phonon energy, are not individually resolved in Fig. 2c. Due to a much stronger outgoing resonance, however, the maximum of the observed phonon resonance is slightly shifted to higher energies as compared to the absorption maximum (in small TPs this shift is more evident, see supplementary Fig. S 1b).

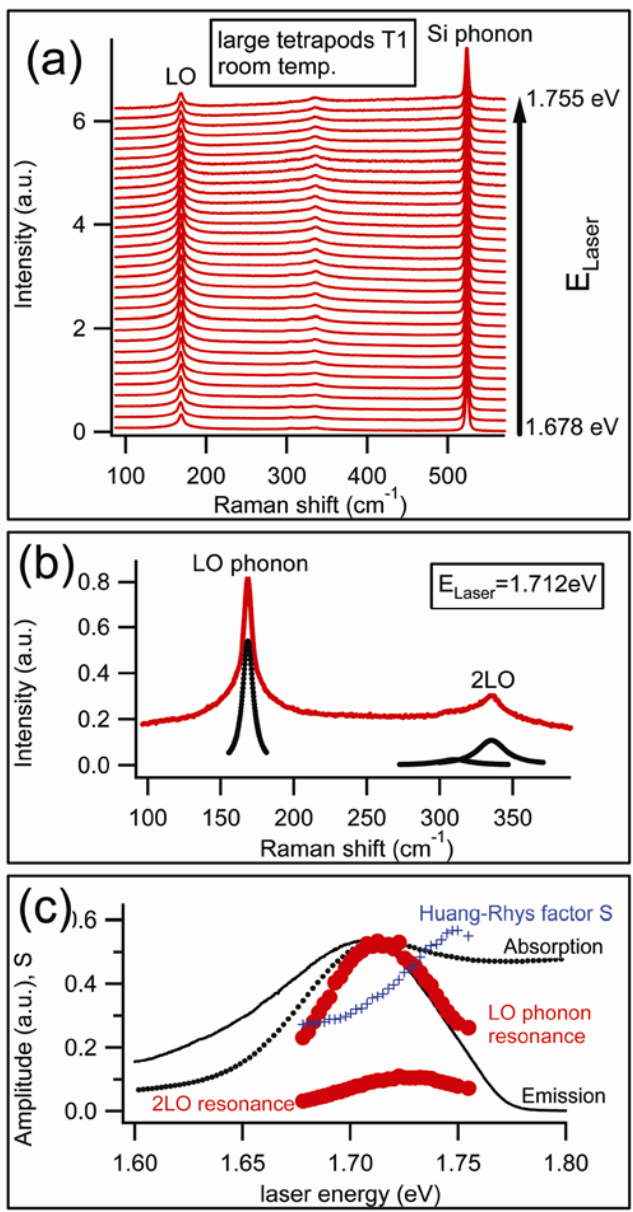

**Figure 2.** Resonant Raman spectra of the large CdTe TPs at room temperature. The TP LO phonon is observed at 168.6 cm$^{-1}$ and its second order scattering at 335.5 cm$^{-1}$. The resonance of the LO phonon with respect to the laser excitation energy is clearly visible. The spectra are normalized with respect to the Si phonon intensity and have been shifted vertically for clarity; (b) Raman spectrum from (a) at resonance maximum. Lorentz fits (black dotted lines) yield as peak positions for LO 168.6 cm$^{-1}$, for 2LO 335.5 cm$^{-1}$, and a third small peak can be identified at 310 cm$^{-1}$;(c) TP LO and 2LO phonon intensity (red thick dots) plotted versus laser energy. The room temperature absorption and emission spectra are shown in black dotted and continuous lines, respectively. Blue markers show the experimental Huang-Rhys factor derived from the ratio of the first and second order scattering of the LO phonon mode.

We note that a double-peak structure is observed in the emission of tetrapods, which becomes more evident in smaller tetrapods, and at low temperatures (see supplementary Fig. S 2b). Theory showed that the high-energy peak in emission can be correlated to transitions from first excited states of electrons and holes that are mainly localized in the tetrapod arms, whereas the low energy peak originates from exciton ground state transitions where the electrons are localized mainly in the tetrapod core[9]. The onset of the LO phonon resonance at low temperatures and in small TPs (see supplementary Fig. S 1 and S 2) occurs clearly at higher energies than the exciton ground state emission and therefore we conclude that the phonon resonance can be correlated to transitions from excited states that are localized in the TP arms. By comparing the resonance maxima of the first and of the second order scattering (at 1.715 eV and 1.730 eV, respectively) we find that the 2LO phonon resonance maximum occurs at a higher laser excitation energy of approximately 15 meV (almost the phonon energy), which further supports the above conclusion that the observed resonances are outgoing resonances.

The Huang-Rhys factor $S$, which is identified with the exciton-phonon coupling in the Franck-Condon model, can be used as a fitting parameter for the relative intensities of different scattering orders. From the data reported in Fig. 2 we can calculate $S$ from the first and the second order scattering amplitudes and we plot its dependence on the exciting laser energy, i.e. on the resonance conditions, shown by the blue markers in Fig. 2c. We find a significant depence of $S$ on the exciting laser energy that results from different maxima positions of the scattering orders discussed above. The absolute values of the Huang-Rhys factor in Fig. 2c are of the same order as those reported by Krauss et al.[29], although they are much larger than those calculated for spherical CdSe nanocrystals[30]. The general understanding is that the Franck-Condon model does not adequately describe the exciton-phonon coupling and that additional, non-adiabatic effects have to be taken into account. In particular, multiphonon scattering has been discussed experimentally and theoretically for spherical nanocrystals by Cardona and coworkers[31] and by Pokatilov et al.[32]. Their calculations show that the discrepancy between experiment and theory can be removed, if band mixing of the exciton states is taken into account, and if additional scattering channels are allowed for the higher order processes. The relative intensities of the different scattering orders in our experiments are very similar to the "strong confinement" condition of Pokatilov et al. [32] which implies also strong band mixing.

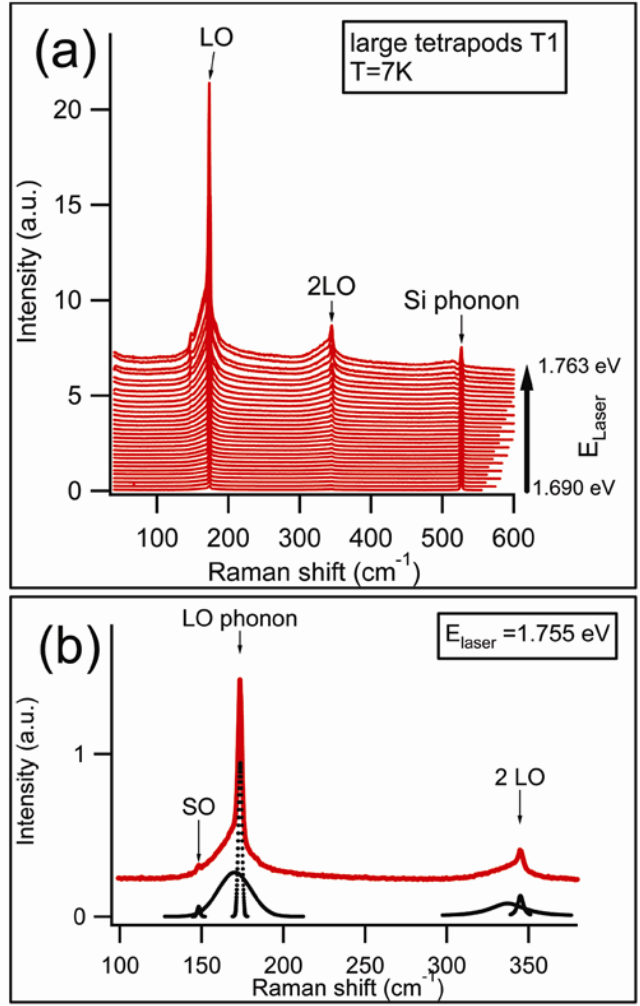

**Figure 3:** (a) Resonant Raman spectra of large CdTe tetrapods at T = 7 K. (b) a spectrum in resonant condition (red line) and Lorentz fits to the data (black dotted lines). We observe a mode at 148 cm$^{-1}$ (FWHM = 1.2 cm$^{-1}$), a mode at 173.5 cm$^{-1}$ (FWHM = 1.5 cm$^{-1}$) and a broad underlying signal that was fitted with a peak at 170 cm$^{-1}$ (FWHM = 14 cm$^{-1}$). Experimental spectra have been shifted for clarity.

In order to resolve the phonon excitations in more detail we performed resonant Raman scattering at low temperatures. Figure 3a shows resonant Raman spectra of large tetrapods (sample T1) at a temperature of T = 7 K. We observe a rich phonon excitation spectrum and find strong resonant enhancement by tuning the laser energy towards the exciton transitions of the tetrapods. We find a small peak at 148 cm$^{-1}$ that becomes clearly observable under strong resonant conditions. A sharp phonon mode is identified at 173.5 cm$^{-1}$ (not broadened significantly with respect to the CdTe bulk LO phonon, see Fig 4), and two higher order scattering orders of this mode are found at 343 cm$^{-1}$ and 515 cm$^{-1}$. For further discussion of the first and second order scattering peaks we plot a spectrum in strong resonance in Fig. 3b. In addition to the above mentioned peaks at 148 cm$^{-1}$, 173.5 cm$^{-1}$, and 343 cm$^{-1}$, Lorentz fits to the data evidence another broad excitation at 170 cm$^{-1}$ and its second order scattering at 337 cm$^{-1}$.

In order to understand the physical origin of the phonon modes we regard the arms of the large tetrapods as nanowires with 12 nm diameter and 80 nm length. Focusing on the mode at 148 cm$^{-1}$ we follow the theory of Gupta and coworkers[16] for surface optical (SO) phonons in nanowires which yields:

$$\omega_{SO}^2 = \omega_{TO}^2 + \frac{\varpi_p^2}{\varepsilon_\infty + \varepsilon_m f(x)}; \quad x = q \cdot r,$$

where $\varpi_p^2 = \varepsilon_\infty \left( \omega_{LO}^2 - \omega_{TO}^2 \right)$ is the screened ion-plasma frequency, $f(x) = \frac{I_0(x) K_1(x)}{I_1(x) K_0(x)}$ (with $I$ and $K$ Bessel functions) and $\varepsilon_\infty$ and $\varepsilon_m$ are the bulk CdTe high frequency and surrounding medium dielectric constants, respectively. For our tetrapod sample T1 we can take the length (l = 80 nm) as the longitudinal symmetry breaking mechanism that leads to $q = 2\pi/l$, r = 6 nm is the arm radius which yields x = 0.47. With $\varepsilon_m$ = 6.5 (average Silicon/vacuum) and $\varepsilon_\infty$ = 7 and $\varepsilon_0$ = 10 we obtain $\omega_{SO}$ = 146 cm$^{-1}$, which is in good agreement with the peak at 148 cm$^{-1}$ in Fig. 3b.

The effect of the nanowire shape on the LO and TO phonons in polar nanocrystals has been investigated by Mahan and coworkers[13]. They predicted a significant splitting of the Raman active TO and LO phonons in large polar nanowires (diameter maximum 5 μm) due to long range dipolar interactions. The bulk TO and LO phonon frequencies are still observed as the z-mode of the phonons and additional modes appear because of the lateral constriction of the wire geometry (z-direction is along, and x-direction is perpendicular to the wire axis). In particular, the predicted values for CdTe wires are: TO$_z$ = TO$_{bulk}$ = 140 cm$^{-1}$, TO$_x$ = 167.3 cm$^{-1}$, LO$_z$ = LO$_{bulk}$ = 171 cm$^{-1}$, and LO$_x$ = 172.4 cm$^{-1}$.

The predicted splitting of the LO phonon is probably too small to be resolved experimentally, in particular if one of the modes is more intense in the spectrum than the other. In general, we would expect the x-confined modes to be broad since they should be sensitive to fluctuations in the nanocrystal size (in arm diameter), and the z-modes to be more sharp due to their bulk-like character. This makes us confident that the sharp phonon excitation at 173.5 cm$^{-1}$ can be attributed to the LO$_Z$ mode and that the broad excitation at 170 cm$^{-1}$ comes from the TO$_X$ phonon mode. The fact that the excitations are observed at slightly higher energies than predicted can be explained by the finite z-confinement.

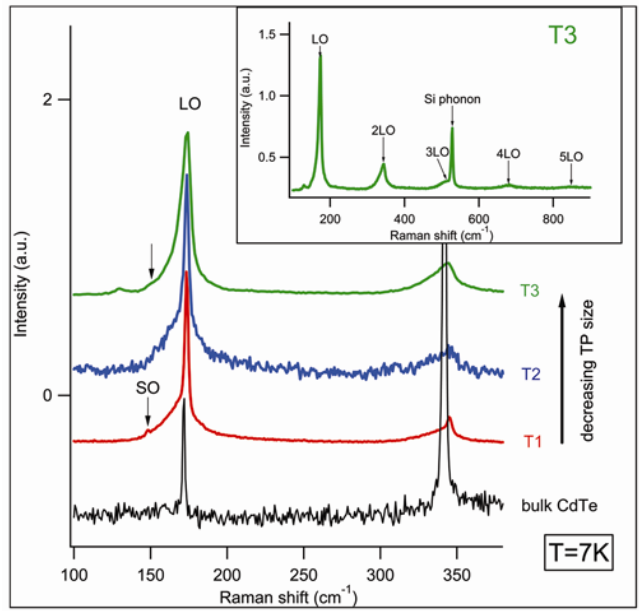

**Figure 4:** Raman spectra of large (red), small (blue), very small (green) tetrapods, and of bulk CdTe (black) at T = 7 K. All TP spectra show the broad peak at the low energy shoulder of the LO$_Z$ phonon excitation. The SO phonon can be clearly identified in the large TPs T1, and can be identified as the small feature indicated by the arrow for T3. As a trend we observe that the intensity of second order scattering increases with decreasing TP dimension. Spectra are normalized with respect to the LO phonon amplitude and have been shifted vertically for clarity. The inset shows a larger range of the T3 spectrum where up to 5 scattering orders are detected. The laser excitation energy for the bulk and samples T1, T2 is 1.755 eV and for sample T3 is 2.33 eV. The excitation at 129 cm$^{-1}$ of sample T3 is due to Te residues[33, 34].

Figure 4 plots the low temperature Raman spectra of large tetrapods (same as in Fig. 3b), of two smaller TP samples T2 and T3 (arm diameter and length 7 nm / 30 nm, and 5 nm / 18 nm, respectively), and of bulk CdTe. The LO phonon of bulk CdTe is detected at 172 cm$^{-1}$, and we find the dominant mode in the TP spectra slightly blue-shifted and slightly broadened at 173.5 cm$^{-1}$. In the experimental spectra, the SO phonon mode is clearly observed in the large TPs T1, and can be weakly noted in the spectrum of T3 (see the arrow). By calculating the SO phonon energy for the small TPs T2 and T3 in the same nanowire framework as for T1, with l = 30 nm, r = 3.5 nm (for T2), and l = 18 nm, r = 2.5 nm (for T3), we obtain SO(T2) = 148 cm$^{-1}$ and SO(T3) = 149 cm$^{-1}$. We note that for sample T1 we observe the SO phonon only in strong resonant conditions (Fig. 3a), and point out that for sample T3 the exciting laser energy of 2.33 eV also meets good vibronic coupling conditions, indicated by the large number of higher order scattering modes (see inset in Fig. 4). However, in sample T2 we do not observe an excitation near 148 cm$^{-1}$, due to the non-resonant conditions. By analyzing the first and second order scattering of TPs with different dimensions we find that, as a trend, the Huang-Rhys factor $S$ increases with decreasing TP size, namely $S$(T1) = 0.17, $S$(T2) = 0.18, and $S$(T3) = 0.23 for the sharp LO phonon. We remark that this conclusion is weakened by the fact that we do not have similar resonance conditions for this comparison. However, we also observe second order scattering for the broad peaks with the same trend: $S_{broad}$(T1) = 0.45, $S_{broad}$(T2) = 0.46, and $S_{broad}$(T3) = 0.62.

In conclusion, we observe strongly resonant phonon excitations that can be assigned to confined optical phonon modes in a nanowire picture. We map in detail the resonance behavior of the phonons with respect to the exciting laser energy and we can correlate the phonon resonances with exciton levels of the tetrapods.

**Acknowledgments:** The authors gratefully acknowledge the support by the SA-NANO European project (Contract No. STRP013698), by the MIUR-FIRB and MIUR 297 (Contract No. 13587) projects and by the German Science Foundation through SFB 508.

**Supporting Information Available**: Resonant Raman Spectra of tetrapods with small arms (sample T2) at room temperature. Resonant Raman Spectra of tetrapods with small arms (sample T2) at T = 7 K.

**Table of Contents graphic:**

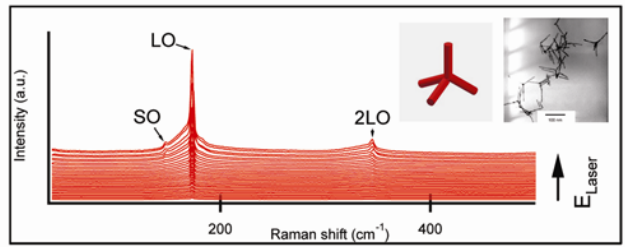